\documentclass[onecolumn,aps,pre,11pt]{revtex4-2}
\usepackage{graphicx} 

\usepackage{amsmath,amsthm,amsfonts,amssymb,mathrsfs}

\usepackage{geometry}

\usepackage{bm}
\usepackage{algpseudocode}
\usepackage{algorithm}

\usepackage{physics}

\usepackage{hyperref}

\usepackage[capitalize]{cleveref}

\def\O{\mathcal{O}}
\def\P{\mathcal{P}}
\def\Q{\mathcal{Q}}

\newtheorem{theorem}{Theorem}
\newtheorem{lemma}{Lemma}

\begin{document}

\title{Exponential Quantum Speedup for Simulating Classical Lattice Dynamics }
\author{Xiantao Li}
\email{Xiantao.Li@psu.edu}
\affiliation{Department  of Mathematics,  \\
The Pennsylvania State University, PA 16802 }

\begin{abstract}
Simulating large-scale lattice dynamics directly is computationally demanding due to the high complexity involved, yet such simulations are crucial for understanding the mechanical and thermal properties of many physical systems. In this work, we introduce a rigorous quantum framework for simulating general harmonic lattice dynamics by reformulating the classical equations as a time-dependent Schr\"odinger equation governed by a sparse Hamiltonian. This transformation allows us to exploit well-established quantum Hamiltonian simulation techniques, offering an exponential speedup with respect to the number of atoms $N$. The overall complexity has a logarithmic dependence on $N$, and linear dependence on both the simulation time $T$ and the Debye frequency $\omega_D$.    Key to our approach is the application of the matrix-valued Fej\'er–Riesz theorem to the phonon dynamical matrix, which facilitates the efficient construction of the underlying Hamiltonian with translational invariance. We demonstrate the applicability of the method across a broad class of lattice models.
\end{abstract}

\maketitle

\section{Introduction}

Lattice dynamics are fundamental to solid-state physics, governing mechanical properties, phonon spectra, elastic wave propagation, and energy transport \cite{maradudin1963theory,peierls1996quantum,AsMe76}. However, direct numerical simulation of these dynamics is intractable for large systems due to the vast number of degrees of freedom. Traditionally, model-reduction techniques—such as memory integrals \cite{adelman1976generalized}, coarse-graining \cite{rudd1998coarse}, continuum approximations \cite{clayton2006atomistic}, and multiscale coupling methods \cite{li2005multiscale,park2005bridging}—have been employed to mitigate this complexity, often at the expense of accuracy.

Quantum computing offers a promising alternative. Recent algorithms that simulate the time-dependent Schr\"odinger equation (TDSE) have demonstrated quantum advantage both theoretically \cite{BCC13,BCK15} and experimentally \cite{clinton2021hamiltonian}. Notably, the recent works \cite{BBK23,CJO19}  have extended these techniques to classical harmonic oscillator networks by encoding harmonic interactions by mapping them to TDSEs through a graph Laplacians and using the incidence matrix (instead of semi-classical limits). However, these methods incur high complexity when negative force constants appear and do not account for vector-valued displacements. 
While the recent works for linear ODEs \cite{PhysRevLett.133.230602,an2023linear,ACL23} also proposed quantum algorithms by reformulating them as TDSEs, the formulations rely on a non-expansive solution property, a condition generally not met by lattice dynamics without resorting to exponential rescaling that diminishes the simulation success probability.

In this paper, we develop a robust quantum algorithm for general \(d\)-dimensional lattice dynamics. Our approach maps the classical dynamics to a TDSE by expressing the force constant matrices through the phonon dynamical matrix. We express this dynamical matrix as a Laurent polynomial on a \(d\)-dimensional tori and apply a matrix-valued Fej\'er–Riesz theorem to obtain a factorization. We show that the resulting factorization yields a TDSE with a sparse Hamiltonian, ensuring efficient quantum simulation. We point out that our method targets classical lattice dynamics—as opposed to quantum lattice models \cite{haah2021quantum}. Overall, this framework offers a practical and scalable pathway for simulating the mechanical and thermal properties of a wide range of physical systems, especially when considered in an ensemble or combined with local defects. Moreover, certain lattice models come from a discretization of elastic wave equations, in which case our method provides an efficient quantum algorithm for simulating wave propagation in elastic media.

\section{ Lattice Dynamics and Quantum Algorithms }

\subsection{From lattice dynamics to TDSE}
A lattice is an infinite collection of points generated by a basis \(\bm{a}_j\) in \(\mathbb{R}^d\):
\begin{equation}
    \mathbb{L} = \left\{ \ell_1 \bm{a}_1 + \ell_2 \bm{a}_2 + \cdots + \ell_d \bm{a}_d \;:\; \ell_1,\ell_2,\dots,\ell_d \in \mathbb{Z} \right\}.
\end{equation}
Complex lattices—formed by superimposing multiple sublattices with relative shifts—can be recast by grouping the displacements within each unit cell onto an equivalent simple lattice isomorphic to \(\mathbb{Z}^d\). Although solid state physics traditionally focuses on three-dimensional (\(d=3\)) and two-dimensional (\(d=2\)) sytems, 
and lattices with even higher dimensions find applications in complex networks \cite{albert2002statistical}.

Following standard notation \cite{AsMe76}, the lattice dynamics are governed by
\begin{equation}\label{eq:newton}
    m_j \frac{d^2}{dt^2}\bm{u}_j = -\sum_{k} D_{j,k}\bm{u}_k,
\end{equation}
where \(\bm{u}_j = \bm{r}_j - \bm{R}_j \in \mathbb{R}^d\) denotes the displacement from equilibrium \(\bm{R}_j\) to the current position \(\bm{r}_j\), and \(D_{j,k}\) is the force constant matrix. For a pair potential \(\phi\), we have \(D_{j,k} = \nabla^2\phi(\bm{R}_j-\bm{R}_k)\); similar definitions hold for multi-body interactions or first-principle calculations. Notably, within the harmonic approximation, \(D_{j,k}\) depends only on the relative positions of the atoms, and thus can be written as $D_{0,k-j}$ as well.
The corresponding Hamiltonian for this classical dynamics is 
\begin{equation}\label{ham}
    H = \frac{1}{2}\sum_j \frac{\bm{p}_j^2}{m_j} + \frac{1}{2} \sum_{j}\sum_k \bm{u}_j^T D_{j,k} \bm{u}_k,
\end{equation}
with \(\bm{p}_j = m_j \dot{\bm{u}}_j\).

Our primary goal is to show that the classical lattice dynamics can be \emph{exactly} mapped onto a time-dependent Schr\"odinger equation (TDSE), allowing us to leverage quantum Hamiltonian simulation algorithms \cite{BCC13,BCK15,gilyen2019quantum}. In matrix-vector form, Eq.~\eqref{eq:newton} reads
\begin{equation}\label{mudu}
    M \frac{d^2}{dt^2}\bm{u} = -D\,\bm{u},
\end{equation}
where \(M\) is a diagonal mass matrix. Following \cite{CJO19,BBK23}, we seek a matrix \(Q\) satisfying
\begin{equation}\label{QTQ=D}
    Q^T Q = D.
\end{equation}
This factorization yields the following equivalence:

\begin{lemma}[\cite{BBK23}]\label{wave2schr}
Suppose there exists a matrix \(Q\) satisfying \(Q^T Q = D\). Then the dynamics of Eq.~\eqref{mudu} are equivalent to the TDSE
\begin{equation}\label{tdse}
    \frac{d}{dt}\ket{\psi(t)} = -iH\,\ket{\psi(t)},
\end{equation}
with
\begin{equation}\label{psi-H}
    \ket{\psi(t)} \equiv \frac{1}{\sqrt{2E}} \begin{bmatrix} M^{1/2}\dot{\bm{u}} \\ i\,Q\,\bm{u} \end{bmatrix}, \quad
    H \equiv -\begin{bmatrix} 0 & M^{-1/2}Q^T \\ Q\,M^{-1/2} & 0 \end{bmatrix}.
\end{equation}
Here $E$ serves as a normalizing parameter, which corresponds to the total energy \eqref{ham}.  
\end{lemma}

\subsection{Decomposition of the matrix $D$}

 In practice, \(Q\) may be non-square. For example, in Ref.~\cite{BBK23} the number of columns in \(Q\) equals the number of edges in the associated graph Laplacian, with entries given by the square roots of the force constants. However, this construction does not readily extend to higher dimensions where the matrices \(D_{j,k}\) are not necessarily symmetric positive definite. Alternatively, one may write Eq.~\eqref{mudu} as a first-order system,
\[
\frac{d}{dt}\bm{u} = M^{-1}\bm{p}, \quad \frac{d}{dt}\bm{p} = -D\,\bm{u},
\]
and apply Schr\"odingerization technique \cite{PhysRevLett.133.230602}, or linear combination of Hamiltonian simulations \cite{an2023linear}, which also reduces the problem to simulating a TDSE. However, direct applications of these methods typically requires a stability condition on the non-Hermitian part that is not generally met by lattice dynamics. One may also use the quantum algorithms for linear differential equations \cite{BCOW17}, in particular, the symplectic integrators \cite{wu2024structure}, which reduce the problem to a quantum linear solver algorithm (QLSA). The  complexity would depend on the condition number of the matrix and solution growth, which may not be available in advance.  

Direct computing the decomposition \eqref{QTQ=D} for large systems is clearly not computationally feasible. 
To find an efficient alternative for which the computation is independent of the system size, we note that \(D_{j,k}\) depends solely on the relative positions \(\bm{R}_j-\bm{R}_k\). We consider finite-ranged interactions—i.e., when \(D_{j,k} = 0\) for \(\|\bm{R}_j-\bm{R}_k\| > r_\mathrm{cut}\). We note that the force constant matrix is related to the dynamical matrix via a Fourier transform:
\begin{equation}\label{Dhat}
    \hat{D}(\bm{\xi}) = \sum_{\substack{\bm{R}_j-\bm{R}_k\in\mathbb{L}\\ \|\bm{R}_j-\bm{R}_k\| < r_\mathrm{cut}}} D_{j,k}\,e^{-i\bm{\xi}\cdot(\bm{R}_j-\bm{R}_k)},
\end{equation}
with \(\bm{\xi}\) in the first Brillouin zone. Expressing \(\bm{\xi} = \theta_1 \bm{b}_1 + \cdots + \theta_d \bm{b}_d\), in the reciprocal basis $\bm b_\beta$ (with \(\bm{a}_\alpha\cdot\bm{b}_\beta=2\pi\delta_{\alpha,\beta}\)), we obtain
\begin{equation}\label{rat-poly}
    \hat{D}(\bm{\xi}) = \sum_{\ell_1,\ldots,\ell_d} D_{\bm{0},\bm{\ell}}\,e^{-i(\theta_1\ell_1+\cdots+\theta_d\ell_d)}.
\end{equation}
Thus, \({D}(\bm{\xi})\) is a matrix-valued, nonnegative rational polynomial defined on the \(d\)-dimensional torus \(\mathbb{T}^d\). We now show that the matrix-valued Fej\'er–Riesz theorem, which produces a factorization of the dynamical matrix $\hat{D}(\bm{\xi})$, provides the desired factorization \eqref{QTQ=D}, providing the matrix \(Q\) required in Lemma~\ref{wave2schr}.

The utility of the Fej\'er–Riesz theorem is particularly transparent in the univariate case. 

\begin{theorem}\label{univariate}
Consider a one-dimensional lattice (\(d=1\)) with $L$ atoms. Let \(D\) be an \(L \times L\) block Toeplitz matrix with block entries \((D)_{j,k}=D_{0,k-j}\in\mathbb{R}^{m\times m}\) for all \(j,k\in [L]\) with \(|j-k|\le p\), $p\in \mathbb{N}$, and \((D)_{j,k}=0\) otherwise. Consider the  Laurent polynomial
\(\P(z) = \sum_{\ell=-p}^{p} D_{0,\ell} \, z^\ell, \quad |z|=1.
\)
There exists an analytic (one-sided) polynomial 
\(\Q(z) = \sum_{\ell=0}^{p} Q_\ell\,z^\ell,\)
such that 
\begin{equation}
    \P(z) = \Q(z)^\dagger \Q(z), \quad |z|=1.
\end{equation}
Defining the \((L+p)\times L\) block matrix \(Q\) with \((j,k)\) block given by \(Q_{j-k}\) for \(
k\leq j \leq k+p \) (and zero elsewhere), then the factorization condition in Eq.~\eqref{QTQ=D} is fulfilled.
\end{theorem}

Here, the degree \(p\) is associated with the cut-off radius in the lattice model. Although the theorem addresses a one-dimensional system, it accommodates \(m\times m\) force constant matrices—relevant, for instance, in diatomic chains where different masses are grouped into a single unit cell, yielding matrix-valued displacements \cite{AsMe76}. We note that the theorem implicitly assumes boundary conditions \(\bm{u}_{0}=\bm{u}_{L+1}=\bm{u}_{-1}=\bm{u}_{L+2}=\cdots=0\). But periodic boundary conditions can be easily incorporated.

\paragraph*{Example: A One-Dimensional Chain with Negative Force Constants.}
We illustrate the theorem with a one-dimensional chain featuring nearest and next-nearest neighbor interactions:
\begin{equation}\label{nnb}
\ddot{u}_j = -\frac{1}{6}\,u_{j-2} + u_{j-1} - \frac{5}{3}\,u_j + u_{j+1} - \frac{1}{6}\,u_{j+2}.
\end{equation}
Here, the negative coefficient \(-\frac{1}{6}\) in the next-nearest neighbor term precludes direct application of methods such as the incident matrix approach in \cite{BBK23}. The corresponding phonon dynamical matrix is given by the Laurent polynomial
\[
\P(z) = -z - \frac{1}{z} + \frac{5}{3} + \frac{1}{6}\,z^2 + \frac{1}{6}\,\frac{1}{z^2},
\]
which, when expressed in terms of \(\theta\) (with \(z=e^{i\theta}\)), leads to the dispersion relation
\[
\omega^2(\theta)=2\,(1-\cos\theta)-\frac{1}{3}\,(1-\cos2\theta).
\]
As \(\P(z)\ge0\), the Fej\'er–Riesz theorem guarantees the existence of an analytic polynomial $\Q(z)=q_0+q_1 z+q_2 z^2, \quad q_j\in\mathbb{R},$
satisfying $\P(z)=|\Q(z)|^2.$

A direct calculation yields
\[
\Q(z)=\frac{1+\frac{1}{\sqrt{3}}}{2}-z+\frac{1-\frac{1}{\sqrt{3}}}{2}\,z^2.
\]
In matrix form, the dynamical matrix \(D\) is a symmetric pentadiagonal \(L\times L\) matrix, and the factor matrix \(Q\) constructed from \(\Q(z)\) is a lower-triangular banded \( (L+2)\times L \) matrix. 
They are explicitly given as:
\[
D = \begin{pmatrix}
\frac{5}{3} & 1 & -\frac{1}{6} & 0 & \dots & 0 & 0 \\
1 & \frac{5}{3} & 1 & -\frac{1}{6} & \dots & 0 & 0 \\
-\frac{1}{6} & 1 & \frac{5}{3} & 1 & \dots & 0 & 0 \\
0 & -\frac{1}{6} & 1 & \frac{5}{3} & \dots & 0 & 0 \\
\vdots & \vdots & \vdots & \vdots & \ddots & \vdots & \vdots \\
0 & 0 & 0 & 0 & \dots & \frac{5}{3} & 1 \\
0 & 0 & 0 & 0 & \dots & 1 & \frac{5}{3}
\end{pmatrix}, \quad
Q = \begin{pmatrix}
q_0 & 0 & 0 & \cdots & 0 & 0 \\
q_1 & q_0 & 0 & \cdots & 0 & 0 \\
q_2 & q_1 & q_0 & \cdots & 0 & 0 \\
0 & q_2 & q_1 & \cdots & 0 & 0 \\
\vdots & \vdots & \vdots & \ddots & \vdots & \vdots \\
0 & 0 & 0 & \cdots & q_1 & q_0 \\
0 & 0 & 0 & \cdots & q_2 & q_1 \\
0 & 0 & 0 & \cdots & 0 & q_2 \\
\end{pmatrix}.
\]
This example clearly demonstrates how the Fej\'er–Riesz factorization enables rigorous treatment of lattice dynamics with both positive and negative force constants. When the lattice dynamics is subject to periodic boundary conditions, the matrix $D$ will have a circulant structure, and the matrix $Q$ can be constructed as a square matrix with a similar circulant structure.

\subsection{Multi-dimensional lattices}

We now return to the multi-dimensional case $d>1$. Due to the phonon stability condition—that is,  \(\hat{D}(\bm{\xi})\) is Hermitian positive semidefinite, and its connection to a Laurent polynomial in \eqref{rat-poly}, 
we can invoke the multivariate Fej\'er–Riesz theorem to arrive at
the factorization \eqref{QTQ=D}.  Unlike the univariate case, the multivariate Fej\'er–Riesz theorem  requires a sum of squares of the form (by writing $\hat{D}$ as $\P$):
\begin{equation}\label{sos}
\P(\bm{z})=\sum_{s=1}^r{\Q}^{(s)}(\bm{z})^\dagger{\Q}^{(s)}(\bm{z}),
\end{equation}
where ${Q}^{(s)}$ are polynomials of degree $q$,
\begin{equation}\label{qsos}
{\Q}^{(s)}(\bm{z})=\sum_{0\le\ell_1,\ldots,\ell_n\leq q} Q^{(s)}_{\bm{\ell}}\,
z_1^{\ell_1 } z_2^{\ell_2 } \cdots z_n^{\ell_n }.
\end{equation}

We generalize \cref{univariate} to a multi-dimensional lattice in a hypercube with $L$ atoms (or unit cells for a complex lattice) in each direction.

\begin{theorem}

Let \( p \in \mathbb{N} \) and consider a \( d \)-dimensional lattice labelled by multi-indices \( {\bm j}=(j_1,\dots,j_n)\in [L]^d \). Suppose \( D \) is an \( L^d \times L^d \) block-Toeplitz matrix, defined by blocks:
\[
(D)_{{\bm j},{\bm k}}=D_{{\bm j},{\bm k}},\quad {\bm j},{\bm k}\in[L]^d,\quad\text{with }D_{\boldsymbol{\bm j, \bm k}}=0\text{ whenever }\|\boldsymbol{\bm j - \bm k}\|_{\infty}>p.
\]
Consider the Laurent polynomial associated with the phonon dynamical matrix,
\[
\P({\bm z})=\sum_{\|\boldsymbol{\ell}\|_{\infty}\leq p}D_{\boldsymbol{\ell}}{\bm z}^{\boldsymbol{\ell}},\quad {\bm z}=(z_1,\dots,z_d)\in\mathbb{T}^d.
\]
Suppose that the generalized multivariate Fej\'er–Riesz theorem applies, that is, there exist analytic polynomials \(\Q^{(s)}({\bm z})\), \( s=1,\dots,r \), of possibly higher degree \( q\ge p \): such that  \cref{sos,qsos} hold. 
Then the matrix $Q$ of dimension \( (L+q)^d \times L^d \times r \)  constructed as follows will satisfy the condition \eqref{QTQ=D}. $\forall \boldsymbol{\ell}\in [L+q]^d$ and ${\bm k}\in [L]^d$.
\begin{equation}\label{Qmat}
    Q_{\boldsymbol{\ell},{\bm k}, s}=Q^{(s)}_{\boldsymbol{\ell}-{\bm k}},\quad\text{if }0\le  (\boldsymbol{\ell}-{\bm k})_j\leq q\text{ for all }j\in [d], \; s\in [r]. 
\end{equation}

\end{theorem}

The main observation is that the force constant matrices are related to the coefficients of $\Q^{(s)}(\bm z)$ as follows
\begin{equation}\label{mat-eq}
    D_{\boldsymbol{0,\ell}}=\sum_{s=1}^{r}\sum_{\substack{{\bm j}\geq{\bm 0}\\ {\bm j}+\boldsymbol{\ell}\geq{\bm 0}\\ \|{\bm j}\|_{\infty},\|{\bm j}+\boldsymbol{\ell}\|_{\infty}\leq q}}(Q^{(s)}_{{\bm j}+\boldsymbol{\ell}})^T Q^{(s)}_{{\bm j}}.
\end{equation}

Here we have only considered a lattice system in a $ d$-dimensional cube. The same result regarding the matrix factorization holds for a general domain. Specifically, let $S \subset \mathbb{Z}^n$ be a finite subset of the $d$-dimensional lattice. The matrix in \cref{mudu}, i.e., $D \in \mathbb{C}^{|S| \times |S|}$, is the matrix with entries $D_{{\bm i},{\bm j}} = D_{{\bm i}-{\bm j}} \quad \text{for all } {\bm i},{\bm j} \in S.$
To construct the matrix $Q$ to fulfill \cref{QTQ=D}, we can construct a slightly larger system by a union with a padded domain,
\[
\widetilde{S} = \bigcup_{{\bm k} \in S} \left\{ {\bm k} + (\alpha_1, \alpha_2, \cdots, \alpha_n) : 0 \le \alpha_1,\alpha_2, \cdots, \alpha_d \le q \right\}.
\]
The padding region includes the immediate neighbors that may or may not lie in $S.$ 
We can construct the matrix $Q \in \mathbb{C}^{r|\widetilde{S}| \times |S|}$, so that $\big(Q\big)_{  \boldsymbol{\ell}, {\bm j},s } =
Q^{(s)}_{\boldsymbol{\ell} - {\bm j}}.$ With direct calculations, one can verify the condition \eqref{QTQ=D} with this construction. 

\subsection{Hamiltonian simulations for the lattice dynamics}

Once such factorizations are obtained, the matrix \( Q \) in \eqref{QTQ=D}, and therefore the Hamiltonian \( H \) in \cref{psi-H}, can be assembled efficiently. Let \( N \) be the total number of atoms; for example, \( N = L^d \) when the system is set up in a \( d \)-dimensional rectangular domain with \( L \) unit cells in each direction.

We first observe that, due to the Toeplitz structure, \( Q \) can be decomposed as
\[
Q = \sum_s \sum_{\bm{j}} A_{\bm{j}}^{(s)} \otimes Q_{\bm{j}}^{(s)},
\]
where \( A_{\bm{j}}^{(s)} \), as indicator matrices for filling in the matrix blocks, have binary elements associated with neighboring atoms, and the number of neighbors in each direction is at most \( 1 \). 

Therefore, our input consists of:  
\textbf{1.} The coefficient matrices \( Q^{(s)}_{\bm{j}} \) of the matrix polynomials \( Q^{(s)}(\bm z) \) from the Fejér–Riesz factorization, whose sizes are independent of the overall lattice size \( N \);  
\textbf{2.} The matrices \( A_{\bm{j}}^{(s)} \), which encode the connectivity of the atoms in the lattice. 

We consider quantum algorithms for the TDSE \eqref{tdse}, and follow \cite{gilyen2019quantum}  to assess the simulation complexity of implementing the dynamics. First, each matrix \( A_{\bm{j}}^{(s)} \) is binary and sparse, with sparsity at most \( 1 \), and can therefore be easily block encoded since $\norm{A_{\bm{j}}^{(s)}}_{\max}= 1$. In particular,  a \( (1, \log N + 3, \epsilon') \) block encoding of \( A_{\bm{j}}^{(s)} \) can be constructed using \( \O\left(\log N + \log^{2.5} \frac{ 1}{\epsilon'}\right) \) one- and two-qubit gates for each $s\in [r], \bm j \in \{0,1,\cdots, q\}^d$.  Secondly, let \( M_\mu \) be the mass matrix for a single unit cell, so that \( M = I_N \otimes M_\mu \). Then in the Hamiltonian from \eqref{psi-H}, the term \( Q M^{-1/2} \) becomes \(\sum_{\bm j,s}A_{\bm{j}}^{(s)} \otimes   Q_{\bm{j}}^{(s)} M_\mu^{-1/2} \). The matrices $ Q_{\bm{j}}^{(s)} M_\mu^{-1/2}$ have dimension $n_A d$ with \( n_A \) being the number of atoms per unit cell. There are at most $r(q+1)^d$ such matrices in total. To incorporate these matrices as block encodings, we  let $\alpha_{\bm j}^{(s)}$ be the corresponding norm.  We use Parseval's identity and show in \cref{phonon-bound} that \( \sum_{\bm j, s} \abs{\alpha_{\bm j}^{(s)}} \)is bounded by \( \alpha_D:=\omega_D n_A d \), where \( \omega_D \) is the Debye frequency— determined by the phonon spectrum. This provides the necessary estimate to block encode $H$ via linear combination of unitaries. The scaling suggests that in the block encoding of  \( A_{\bm{j}}^{(s)} \) and $ Q_{\bm{j}}^{(s)} M_\mu^{-1/2}$, we set $\epsilon'= \frac{\epsilon}{\alpha_D T} $ for a simulation time $T$.

With the structure of the Hamiltonian \( H \) in \cref{psi-H} established, and by invoking the optimal Hamiltonian simulation algorithm \cite{gilyen2019quantum}, we arrive at the following estimate for the simulation complexity.

\begin{theorem}
Consider the lattice dynamics \eqref{eq:newton} under either periodic or homogeneous boundary conditions. If the phonon dynamical matrix \( \hat{D} \) admits a Fejér–Riesz factorization \eqref{sos}, then there exists a quantum algorithm that simulates the dynamics by producing the quantum state \( \ket{\psi(T)} \) at time \( t = T \), which encodes the displacement and velocity as in \eqref{psi-H}. Given precision \( \epsilon \), the algorithm requires
\[
\mathcal{O}\left( \alpha_D T + \log \frac{1}{\epsilon} \right)
\]
queries to the matrices \( A_{\bm{j}}^{(s)} \) and \( Q_{\bm{j}}^{(s)} M_\mu^{-1/2} \), using
\[
\O\left( r(q+1)^d \left(\log N  +\log^{2.5} \frac{ \alpha_D T }{\epsilon}  \right)^2\left( \alpha_D T + \log \frac{1}{\epsilon} \right) \right)
\]
one- and two-qubit gates.  
Here the parameter \( \alpha_D = \omega_D n_A d  \) is a constant that depends on the phonon spectrum.
\end{theorem}

In the theorem above, \( d \) denotes the spatial dimension of the lattice dynamics. For applications in solid-state physics, the most relevant cases are \( d = 2,3 \). The factorization assumption \eqref{sos} is known to hold for \( d \leq 2 \), with the two-dimensional case recently established in \cite{dritschel2025factoring}. Moreover, the results in \cite{dritschel2004factorization} showed that such factorizable polynomials are dense. In this work, we employ numerical techniques to test factorizability and compute the polynomials \( \Q^{(s)} \) for various lattice structures across \( d = 1,2,3 \), covering both simple and complex lattices. We also find in all our numerical tests that $r\leq 2$ and $q=p$.   Notably, determining factorizability is independent of the number of atoms and therefore is separate from the  complexity of the quantum algorithm.

The quantum state \( \ket{\psi(T)} \) automatically encodes the velocity. To recover the displacement, we exploit the lower-triangular structure of the matrix \( Q \) in \eqref{Qmat}. This allows the displacement to be obtained via forward substitution, involving only \( O(q^d) \) terms. This procedure is valid provided that \( Q_0^{(s)} \), for some \( s \in [r] \), has full rank—a condition we have verified to hold in all of our numerical simulations. With the encoding into \( \ket{\psi(T)} \), many other physical properties, including local energy and density of states, can also be estimated by recasting them as expectation values \cite{BBK23,liu2024toward}.

\section{Examples}

We present numerical results for several lattice structures to illustrate the role of Fej\'er–Riesz factorization in mapping classical lattice dynamics to a TDSE. We demonstrate that the required factorization can be obtained either analytically, or numerically through optimization techniques. We also showcase examples of lattice dynamics simulations that can be efficiently carried out by solving the corresponding TDSEs. The source code used to generate these results is available in \cite{Li-Lattice2025}.

\subsection{One-dimensional lattice with nearest and next nearest neighbor interactions}

We conduct a numerical experiment demonstrating the propagation and reflection of a Gaussian wave packet in a one-dimensional lattice system \eqref{nnb} comprising \(L = 127\) atoms. The dynamics are governed by interactions with nearest and next-nearest neighbors \eqref{nnb}. The atoms' initial displacement and velocity are given by a Gaussian wave packet localized around atom \(x_0 = L/4\) with a Gaussian envelope modulated by a cosine function of wave number \(k_0 = 1.2\):
\[
u_j(0) = e^{-\frac{(j - x_0)^2}{2 \sigma^2}}\cos(k_0 j),\quad v_j(0) = -\omega_{k_0}\, e^{-\frac{(j - x_0)^2}{2 \sigma^2}}\cos(k_0 j),
\]
where the width parameter is set as \(\sigma = 6\).
The lattice dynamics follow Newton's equations \eqref{nnb} with a pentadiagonal stiffness matrix \(D\) capturing nearest and next-nearest neighbor interactions. 

\begin{figure}[thpb]
    \centering
    \includegraphics[scale=0.3]{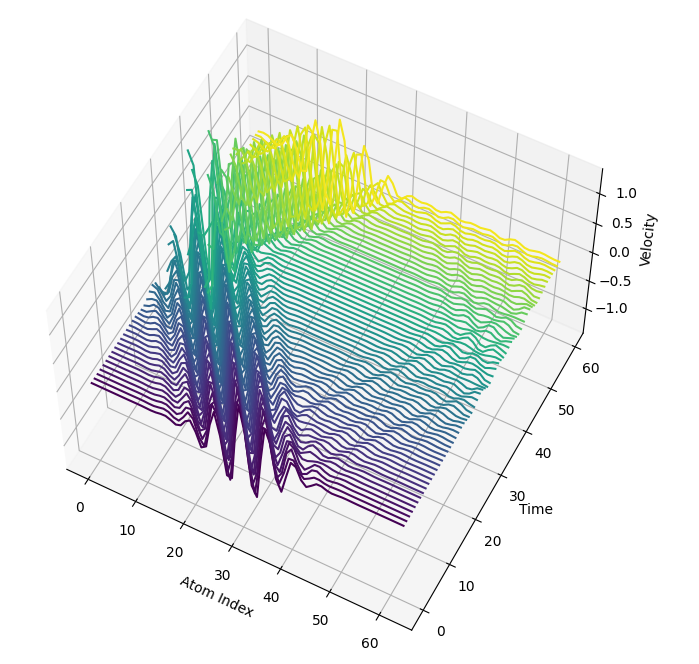}
    \includegraphics[scale=0.3]{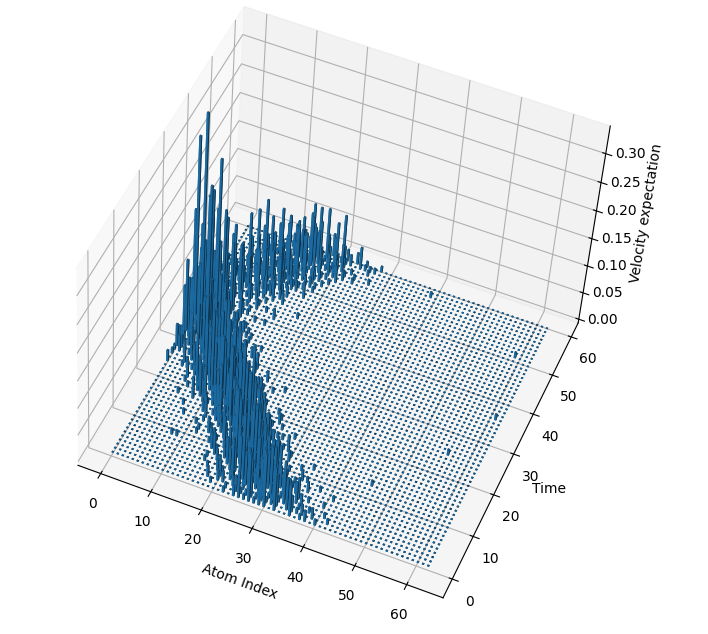}
    \caption{Simulation of a wave packet. Left: direct simulation of the lattice dynamics \cref{nnb} using Verlet's scheme. Right: simulation via the corresponding TDSE \eqref{tdse}  using a quantum simulator.  }
    \label{fig:1dnnb}
\end{figure}

The system's dynamics are numerically integrated using Verlet's integration scheme with time step \(\Delta t = 0.05\). The total simulation time is \(T = 60\). Figure \ref{fig:1dnnb} (left panel) shows the snapshots of the velocity at different simulation times.

Next, we simulate the same lattice dynamics by mapping classical wave packet propagation into a quantum framework using Qiskit and the embedding into the TDSE in \cref{psi-H,tdse}.  The TDSE governing the state evolution is solved using Trotter decomposition, implemented as quantum gate operations within Qiskit.  
Measurement operators corresponding to individual velocity components are performed with 100 shots per measurement.  The estimated velocity is shown in the right panel of \cref{fig:1dnnb}, and the agreement with the previous results confirms the feasibility of the approach using \cref{psi-H} and \cref{univariate}.

\subsection{One-dimensional Diatomic Chain}

Another classic model in solid-state physics is the one-dimensional diatomic chain, where atoms alternate between two different masses. In our model, atoms at odd-numbered sites have mass $m_A$, and those at even-numbered sites have mass $m_B$. The equations of motion are given by
\[
m_j \ddot{u}_j = u_{j-1} - 2u_j + u_{j+1}.
\]

Let us illustrate the use of unit cells with this example. We define   a two-atom unit cell via the displacement vector $\bm{u}_j = \begin{pmatrix} u_{2j-1}, \; u_{2j} \end{pmatrix}, $ then
the equations of motion become
\[
\begin{pmatrix} m_A & 0 \\ 0 & m_B \end{pmatrix} \ddot{\bm{u}}_j = D_{-1}\bm{u}_{j-1} + D_0 \bm{u}_j + D_1 \bm{u}_{j+1},
\]
with the force constant matrices given by,
\[
D_{-1} = \begin{pmatrix} 0 & -1 \\[1mm] 0 & 0 \end{pmatrix}, \quad
D_0 = \begin{pmatrix} 2 & -1 \\[1mm] -1 & 2 \end{pmatrix}, \quad
D_1 = D_{-1}^T.
\]

The dynamical matrix may be represented as a matrix-valued Laurent polynomial:
\[
{\P}(z) = \frac{1}{z}\begin{pmatrix}0 & -1\\[1mm]0 & 0\end{pmatrix} + z\begin{pmatrix}0 & 0\\[1mm]-1 & 0\end{pmatrix} + \begin{pmatrix}2 & -1\\[1mm]-1 & 2\end{pmatrix}, \quad z=e^{i\theta}.
\]

By the matrix-valued Fej\'er–Riesz theorem, there exists an analytic polynomial matrix $\Q(z)$ of degree one such that $\P(z)= \Q(z)^* \Q(z),\quad |z|=1,$ 
with $\Q(z)= Q_0 + Q_1 z,$
where a suitable choice is
\[
Q_0 = \begin{pmatrix} 1 & 0 \\[1mm] 1 & -1 \end{pmatrix}, \quad Q_1 = \begin{pmatrix} 0 & -1 \\[1mm] 0 & 0 \end{pmatrix}.
\]

In block matrix form, the full dynamical matrix is written as
\[
D = \begin{pmatrix}
D_0 & D_1 & 0 & \cdots & 0 \\[3mm]
D_{-1} & D_0 & D_1 & \cdots & 0 \\[3mm]
0 & D_{-1} & D_0 & \ddots & \vdots \\[3mm]
\vdots & \vdots & \ddots & \ddots & D_1 \\[3mm]
0 & 0 & \cdots & D_{-1} & D_0
\end{pmatrix}_{2L\times 2L},
\]
and the corresponding factorization matrix is
\[
Q =
\begin{pmatrix}
Q_0 & 0 & \cdots & 0 \\
Q_1 & Q_0 & \cdots & 0 \\
0 & Q_1 & \ddots & \vdots \\
\vdots & \vdots & \ddots & Q_0 \\
0 & 0 & \cdots & Q_1
\end{pmatrix}_{2(L+1) \times 2L}.
\]
We show these matrices to provide a more intuitive view of the matrix $Q$, constructed from the matrix factor $\Q(z).$

We employ the above factorization to evolve the system via the associated Schr\"odinger equation \eqref{psi-H} and perform ensemble simulations. Specifically, we consider a diatomic chain of $127$ atoms with alternating masses $m_A=1$ and $m_B=1.5$. The initial velocities are chosen as independent Gaussian random variables with variances that depend on the atom's position, reflecting a spatially inhomogeneous kinetic energy distribution. In our simulations, a Gaussian profile with a peak value of $2.0$, centered at the 12th atom, is used to initialize the system. We perform $1024$ realizations and compute the local kinetic energy for each atom. 

\begin{figure}
    \centering
    \includegraphics[scale=0.5]{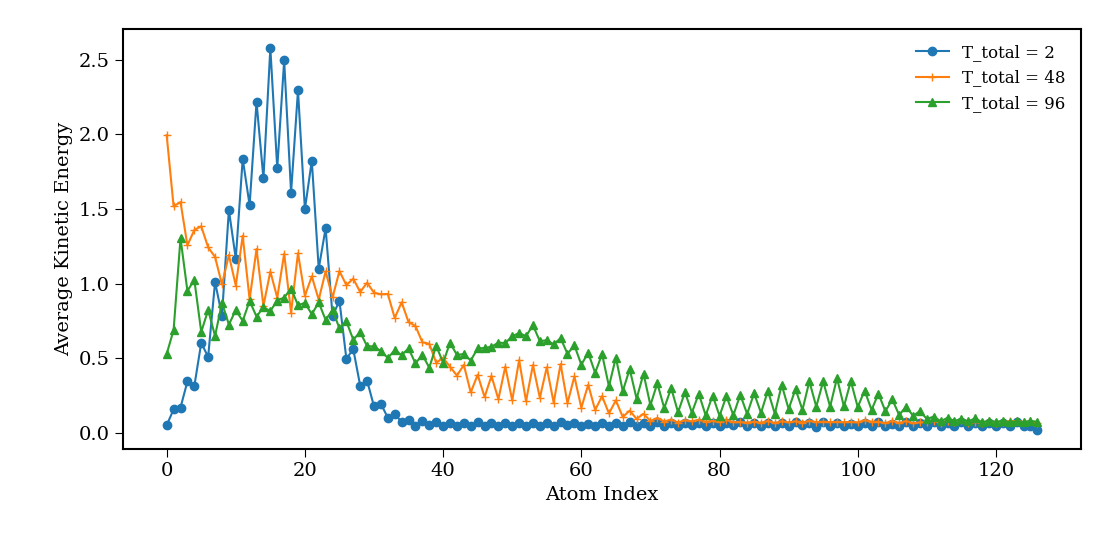}
    \caption{Snapshots of the local kinetic energy in a diatomic chain at several time instances, obtained from ensemble simulations of the corresponding Schr\"odinger equation \eqref{tdse}. }
    \label{fig:ensemble-md}
\end{figure}

Figure~\ref{fig:ensemble-md} shows snapshots of the local kinetic energy at several time instances, illustrating how the profile of the local kinetic energy evolves over time. The observed dynamics are indicative of heat conduction in low-dimensional systems, a phenomenon that has been studied extensively in previous works \cite{lepri2003thermal}. Our results, obtained from Schr\"odinger-equation-based ensemble simulations, provide a new approach for studying thermal transport mechanisms in lattice systems.

\subsection{Two-dimensional Graphene Lattice}

\begin{figure}[tbhp]
    \centering
    \includegraphics[width=0.5\linewidth]{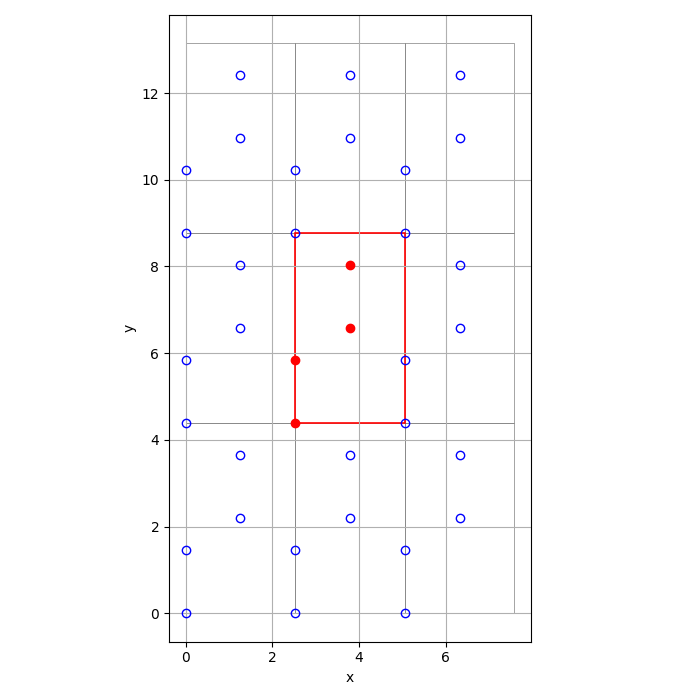}
    \caption{Graphene lattice. The four atoms in the unit cell are highlighted.}
    \label{fig:graph}
\end{figure}

We now consider a two-dimensional system: the graphene lattice. \Cref{fig:graph} illustrates the crystal structure of graphene, constructed from a rectangular supercell containing four carbon atoms per unit cell. The full lattice is generated by translating this unit cell along two orthogonal lattice vectors, resulting in a periodic tiling of the graphene sheet. In the orientation shown, the zigzag direction lies along the horizontal axis, while the armchair direction is vertical.

Relative to the central unit cell in \Cref{fig:graph}, there are eight neighboring cells. The force constants are computed using the Tersoff potential \cite{tersoff1988empirical}. Based on the relative positions of these neighbors, we label the corresponding $8\times8$ force constant matrices as $D_{i,j}$, where $i,j = -1,0,1$. This setup leads to a matrix polynomial of degree $p=2$ , owing to interactions with neighbors at offsets such as $(-1,-1)$ and $(1,1)$. We seek analytic factorizations of this matrix polynomial, also of degree $q=2$, via a nonlinear least squares approach that directly enforces the matrix equations \eqref{mat-eq}. \Cref{tab:optimization_error} reports the optimization error for several choices of the rank parameter $r$. When $r=1$, the error remains sizable, indicating that a factorization is not feasible; however, for $r\geq2$, the error becomes negligibly small, suggesting that an analytic factorization exists.

\begin{table}[htbp]
  \centering
  \begin{tabular}{|c|c|}
    \hline
    \textbf{Parameter $r$} & \textbf{Optimization Error} \\
    \hline
    1 & $1.51543\times10^{-1}$ \\
    2 & $1.76253\times10^{-7}$ \\
    3 & $1.75851\times10^{-7}$ \\
    \hline
  \end{tabular}
  \caption{Optimization error for different values of the rank parameter $r$.}
  \label{tab:optimization_error}
\end{table}

We next perform numerical simulations of lattice dynamics in a graphene sheet containing a square vacancy—a geometry that has been extensively studied (e.g., \cite{cupo2017periodic}). In our simulation, a localized wave packet is constructed to propagate toward the vacancy. The initial wave packet is prepared by modulating the velocity field to generate a well-defined group velocity directed toward the void. The dynamics are simulated by evolving the time-dependent Schr\"odinger equation \eqref{tdse}, with the Hamiltonian constructed using the \(Q\) factorization described above.

\begin{figure}[htpb]
    \centering
    \includegraphics[width=0.3\linewidth]{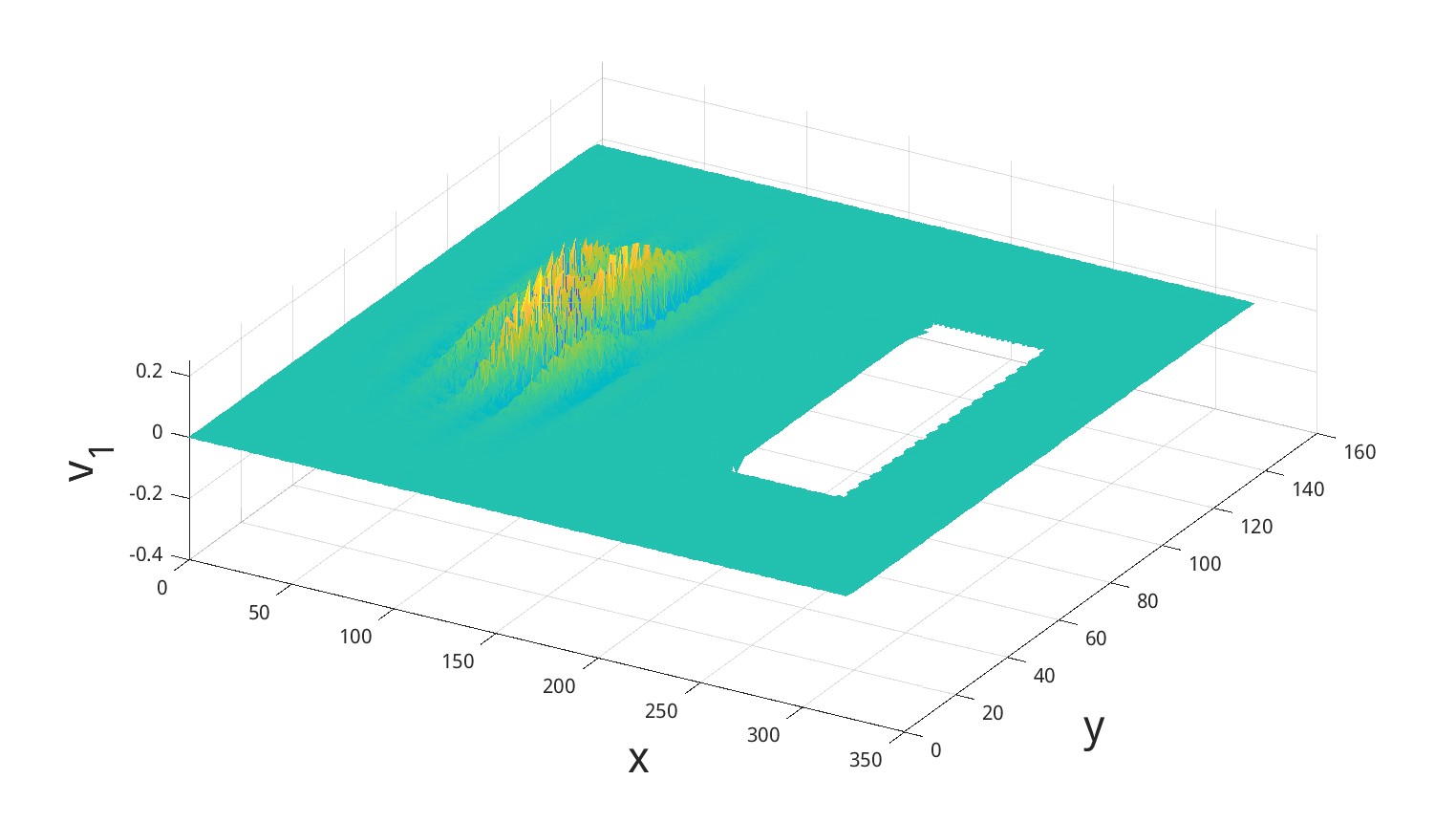}
    \includegraphics[width=0.3\linewidth]{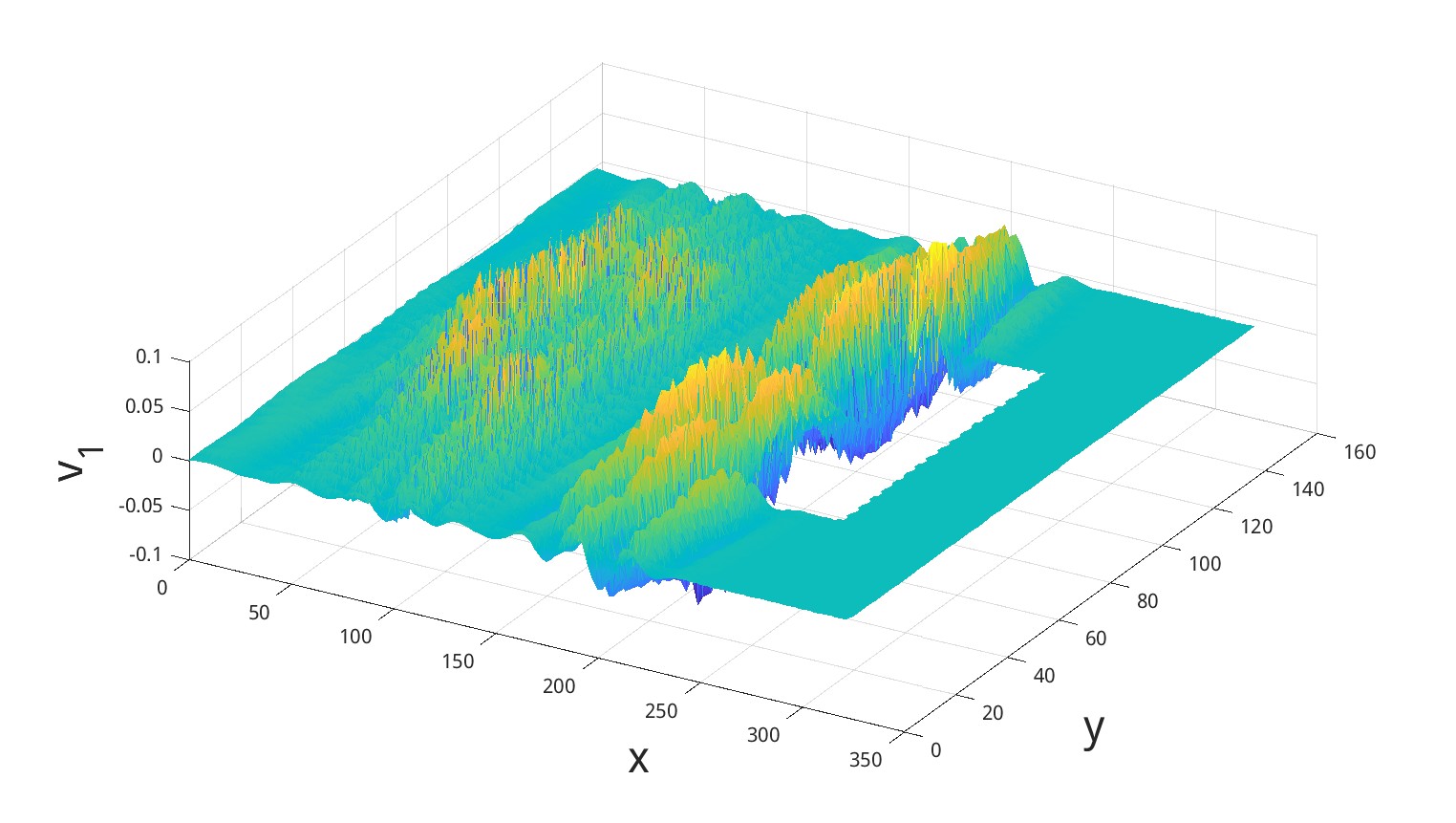}
    \includegraphics[width=0.3\linewidth]{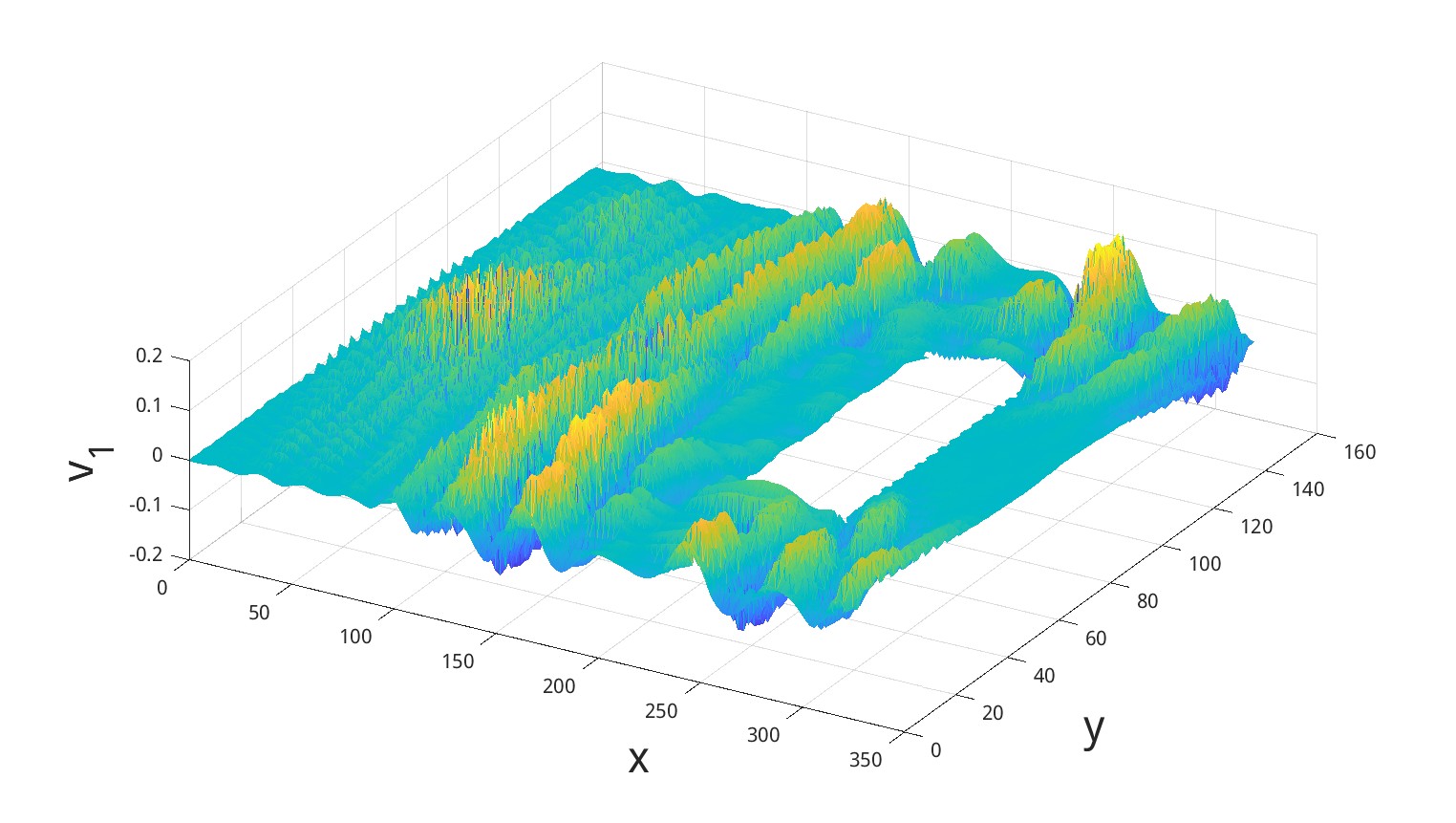}
    \caption{Simulation of a lattice wave in a graphene sheet with a square vacancy. The system is built from $128\times 32$ unit cells.  Left: The first component of the velocity field at $T=0$. Middle: The wave packet arrives at the vacancy at $T=6.144\,\mathrm{ps}$. Right: The velocity profile after interaction with the vacancy at $T=12.288\,\mathrm{ps}$, illustrating scattering effects.}
    \label{fig:gwave}
\end{figure}

Figure~\ref{fig:gwave} shows the first component of the velocity field.     The simulation, based on the time-dependent Schr\"odinger equation \eqref{tdse} derived from the \(Q\) factorization, captures defect-induced dynamical phenomena in the 2D lattice. These results demonstrate the efficacy of our algorithm in 2D systems, especially in capturing defect-induced scattering phenomena.

\subsection{Three-Dimensional Face-Centered Cubic Lattice}

\begin{figure}[htbp]
    \centering
    \includegraphics[scale=0.35]{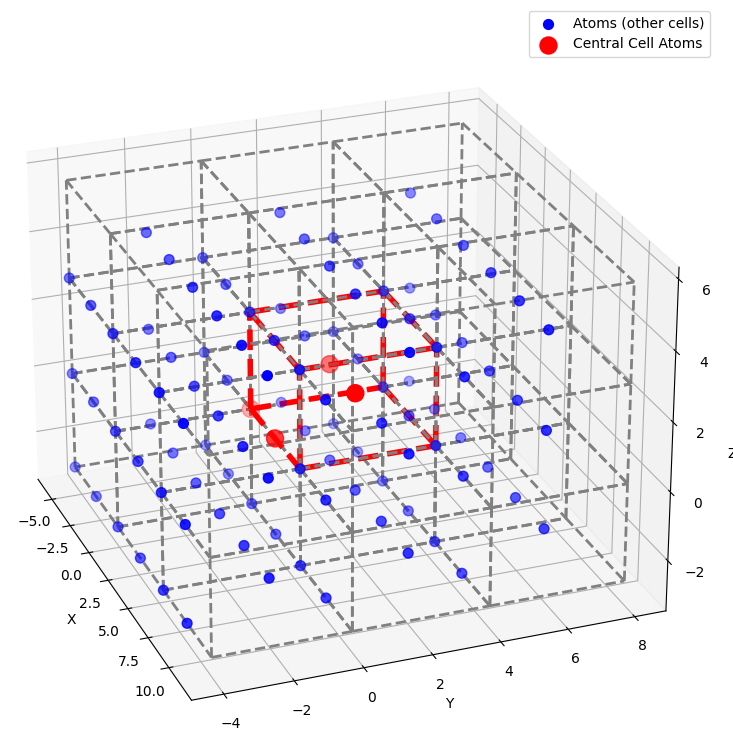}
    \caption{Supercell of an FCC lattice with periodic images. The central cell (red) contains the primitive FCC unit cell, while the 26 surrounding cells (blue) illustrate the lattice periodicity.}
    \label{fig:fcc}
\end{figure}

We now extend our analysis to a three-dimensional lattice. As described above, matrix polynomials can be constructed either using a reciprocal lattice basis or by defining a supercell with an orthogonal basis. Here, we focus on a face-centered cubic (FCC) lattice for aluminum. Specifically, we choose a unit cell oriented along the crystallographic directions \([110]\), \([001]\), and \([1\bar{1}0]\). As shown in \Cref{fig:fcc}, each unit cell contains four atoms, so that the corresponding force constant matrices are \(12\times 12\). In our numerical experiments, these matrices are computed using an embedded atom potential \cite{ercolessi1994interatomic}. The unit cell dimensions are given by $\Delta x = \sqrt{2}\,a_0,\quad \Delta y = a_0,\quad \Delta z = \frac{\sqrt{2}}{2}\,a_0,$
with \(a_0=4.032\,\text{\AA}\) being the lattice constant. 

\begin{figure}[htbp]
    \centering
    \includegraphics[scale=0.35]{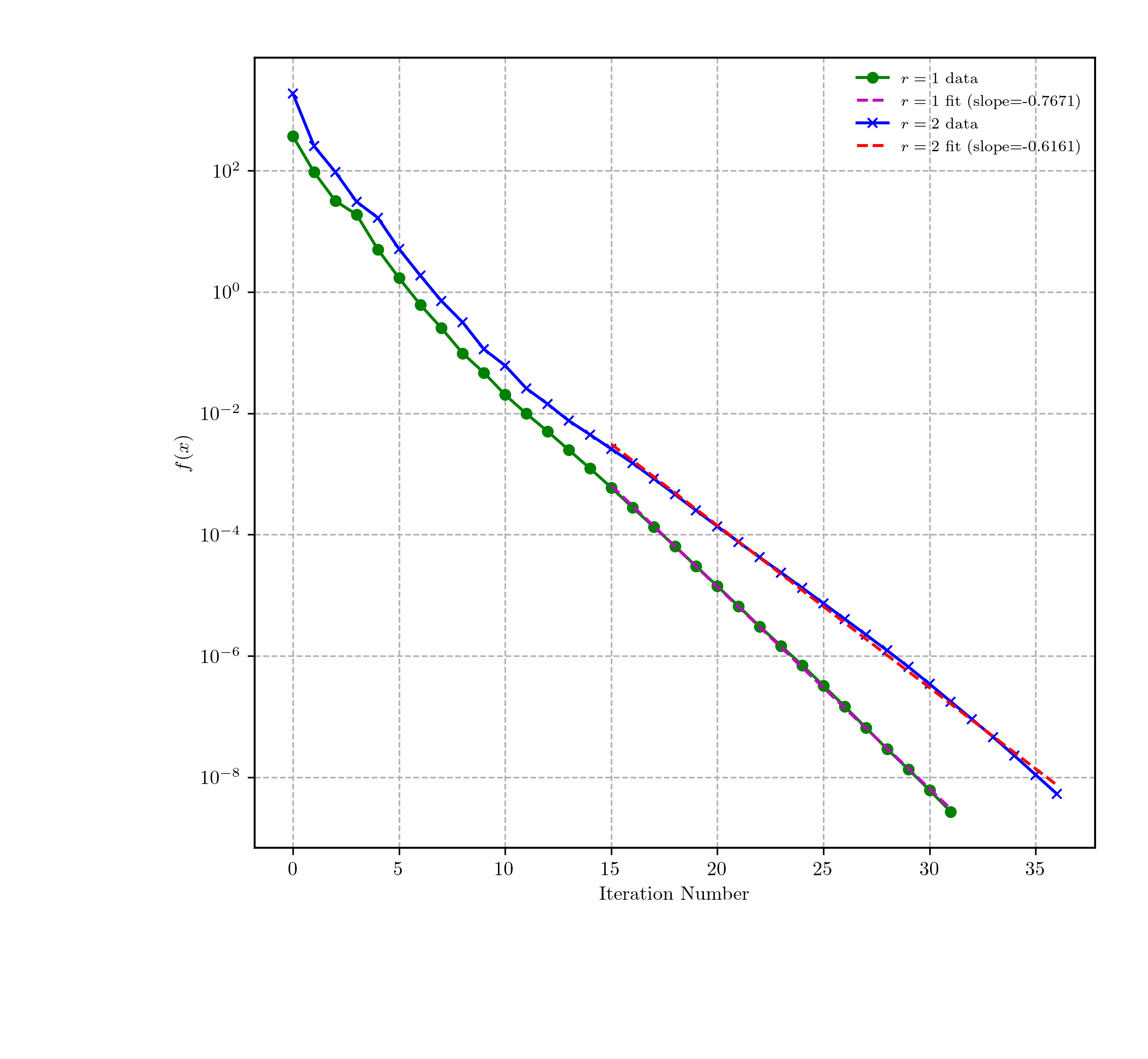}
    \caption{Optimization error for the matrix factorization with \(r=1\) (green) and \(r=2\) (blue). Exponential convergence is observed in both cases.}
    \label{fig:optimization3d}
\end{figure}

Using the force constant matrices and the supercell structure illustrated in \Cref{fig:fcc}, we perform the matrix factorization via an optimization approach. A nonlinear least-squares loss function is defined to minimize the residual error in the matrix equation \eqref{mat-eq}, with the error measured in the Frobenius norm. We consider both the \(r=1\) and \(r=2\) cases. \Cref{fig:optimization3d} displays the optimization error versus iteration number until convergence. Exponential convergence is observed in both cases, suggesting that such factorization exists even for $r=1$.

\section{Discussions}

In this work, we have established a rigorous and exact mapping from classical lattice dynamics to time-dependent Schr\"odinger equations with sparse Hamiltonians, enabling efficient quantum simulation of such systems.  The recent implementation \cite{luangsirapornchai2025practical} of quantum algorithms for harmonic oscillators suggests the possible near-term feasibility of this type of approach. Furthermore, the connection we have built between lattice dynamics and TDSEs opens the door to applying quantum simulation techniques to problems in solid-state physics that involve lattice dynamics as a mechanical or heat reservoir. 

This framework suggests several directions for future work. It may serve as a foundation for tackling broader classes of problems related to lattice dynamics, including statistical ensembles and systems with external forces or boundary effects. The approach naturally extends to incorporate lattice imperfections such as defects, by interfacing with a nonlinear defect model, thus making it applicable to realistic material systems where phonon-defect interactions play a central role. In addition, by interpreting lattice models as discretizations of elastic wave equations \cite{clayton2006atomistic,CJO19}, the method provides a pathway toward quantum algorithms for solving continuum wave phenomena. Finally, the framework remains applicable to systems with long-range interactions, though such cases may require the development of new techniques to address the challenges posed by non-sparse Hamiltonians, e.g., \cite{wang2020quantum}.

\paragraph*{Acknowledgement.} This research is supported by the NSF Grant DMS-2411120.


%

\appendix

\section{Proof of the factorization for the one-dimensional case } \label{thm2-proof}

\begin{proof}
Given the Laurent polynomial \(\P(z)\), the Fej\'er–Riesz theorem guarantees the existence of a polynomial \(\Q(z)\) \cite[Theorem 1.1]{dritschel2004factorization} such that
\[
\P(z) = \Q(z)^\dagger \Q(z), \quad \Q(z)=\sum_{\ell=0}^{p} Q_\ell\,z^\ell,\quad z\in \mathbb{T}.
\]
Equating coefficients yields
\begin{equation}\label{D2Q}
D_{0,\ell}=\sum_{j=0}^{p-\ell} Q_{j+\ell}^T Q_{j}, \quad \ell\ge 0; \quad D_{0,\ell}=D_{0,-\ell}^\dagger, \quad \ell<0.
\end{equation}
By examining the block matrix product 
  \[(Q^TQ)_{j,k} = \sum_{\ell=1}^{L+p} \left(Q\right)_{\ell,j}^T \left(Q\right)_{\ell,k} 
  =  \sum_{i=0}^{p} Q_{i+k-j}^T Q_{i} = D_{0,k-j} = D_{j,k}, 
  \]
  we see that the condition \eqref{QTQ=D} holds. 
\end{proof}

\section{Proof of the bounds on the coefficients in $\Q(z)$}\label{phonon-bound}

\begin{proof}
Fix any $\ell \in [r] $ and write
\[
\Q^{(\ell)}({\bm z}) = \sum_{{\bm j} } Q^{(\ell)}_{{\bm j}} {\bm z}^{{\bm j}}.
\]
By Parseval’s identity over the torus $\mathbb{T}^d$ and unitarity of the Fourier basis, we have
\[
\sum_{{\bm j} } \|Q^{(\ell)}_{{\bm j}}\|_F^2 \leq \int_{\mathbb{T}^d} \|\Q^{(\ell)}({\bm z})\|_F^2 \, d\mu({\bm z})
\leq \int_{\mathbb{T}^d} \|\P({\bm z})\|_F^2 \, d\mu({\bm z})\leq m \cdot \|\P({\bm z})\|_\infty.
,
\]
where $\mu$ is the normalized Haar measure on $\mathbb{T}^d$. By including the mass matrix $M_\mu$, we get the bound by noticing that,
\[
\|{M^{-1/2}_\mu \P({\bm z}) M^{-1/2}_\mu} \| \leq \omega_D.
\]

\end{proof}

\end{document}